\documentclass[aps,prl,twocolumn,preprintnumbers,nofootinbib,10pt]{revtex4-1}%

\pdfoutput=1
\usepackage{amsmath,amsthm,amssymb,multirow,psfrag}
\usepackage{epsfig}
\usepackage{color}

\graphicspath{{./Figures/}}

\begin{document}

\def\lsim{\mathrel{\rlap{\lower4pt\hbox{\hskip1pt$\sim$}}F
  \raise1pt\hbox{$<$}}}
\def\gsim{\mathrel{\rlap{\lower4pt\hbox{\hskip1pt$\sim$}}
  \raise1pt\hbox{$>$}}}
\newcommand{\vev}[1]{ \left\langle {#1} \right\rangle }
\newcommand{\bra}[1]{ \langle {#1} | }
\newcommand{\ket}[1]{ | {#1} \rangle }
\newcommand{\ev}{ {\rm eV} }
\newcommand{\kev}{{\rm keV}}
\newcommand{\mev}{{\rm MeV}}
\newcommand{\tev}{{\rm TeV}}
\newcommand{\mpl}{$M_{Pl}$}
\newcommand{\mw}{$M_{W}$}
\newcommand{\Ft}{F_{T}}
\newcommand{\Zparity}{\mathbb{Z}_2}
\newcommand{\BLambda}{\boldsymbol{\lambda}}
\newcommand{\met}{\;\not\!\!\!{E}_T}
\newcommand{\suRL}{$M_{W}$}

\newcommand{\beq}{\begin{equation}}
\newcommand{\eeq}{\end{equation}}
\newcommand{\bea}{\begin{eqnarray}}
\newcommand{\eea}{\end{eqnarray}}
\newcommand{\nn}{\nonumber \\ }
\newcommand{\gev}{{\mathrm GeV}}
\newcommand{\hc}{\mathrm{h.c.}}
\newcommand{\eps}{\epsilon}

\newcommand{\cO}{{\cal O}}
\newcommand{\cL}{{\cal L}}
\newcommand{\cM}{{\cal M}}

\newcommand{\Deltab}{ {{\bar{\Delta}}} }
\newcommand{\Hb}{ {{\bar{H}}} }
\newcommand{\Phb}{ {{\bar{\Phi}}} }
\newcommand{\Tr}[1]{\ensuremath{{\bf Tr}[\,#1\,]}} 

\newcommand{\Chi}{ \ensuremath{X} }
\newcommand{\Phib}{ \ensuremath{\pmb{\Phi}} }
\newcommand{\Chid}{\Chi^\dagger}
\newcommand{\Phid}{\Phi^\dagger}
\newcommand{\Chic}{\Chi_c}
\newcommand{\Phic}{\Phi_c}
\newcommand{\Chicstar}{\Chi_c^\dagger}
\newcommand{\Phicstar}{\Phi_c^\dagger}
\newcommand{\txx}{\ensuremath{{\Tr{ \Phic \sigma_i \Phi \sigma_j }  \left[ V \Chi^\dagger V^\dagger \right]_{i,j}  + c.c.} }}
\newcommand{\txcO}{\ensuremath{{\Tr{\Chi^\dagger\Chi\Chi^\dagger\Chi} - \Tr{\Chi^\dagger\Chi}^2}}}
\newcommand{\txcTc}{\ensuremath{{\Tr{ \Chid T_i \Chicstar T_j}\, \Tr{ \Phi \sigma_j \Phic\sigma_i} + c.c.}}}
\newcommand{\txcEc}{\ensuremath{{\Tr{ \Chid T_i \Chi T_j }\, \left[ V \Chi V^\dagger \right]_{i,j} + c.c.}}}
\newcommand{\txcFc}{\ensuremath{{\Tr{ \Phid \sigma_i \Phi \sigma_j}\, \left[ V \Chi V^\dagger \right]_{i,j}  + c.c.}}}
\newcommand{\txcIf}{\ensuremath{{\Tr{ \Chid \Chi} \, \Tr{ \Phid \Phi} }}}
\newcommand{\txcIh}{\ensuremath{{ \Tr{ \Chid T_i \Chi}\, \Tr{ \Phid \sigma_i \Phi}}}}
\newcommand{\txcIcO}{\ensuremath{{ \Tr{ \Chicstar  T_i \Chi_c}\, \Tr{ \Phicstar \sigma_i \Phic   } }}}
\newcommand{\txcIc}{\ensuremath{{ \Tr{ \Chid  T_i \Chi T_j }\,  \Tr{ \Phid\sigma_i \Phi \sigma_j }}}}
%
%

\graphicspath{{./Figures/}}

\newcommand{\fref}[1]{Fig.\,\ref{fig:#1}} 
\newcommand{\eref}[1]{Eq.~\eqref{eq:#1}} 
\newcommand{\aref}[1]{Appendix~\ref{app:#1}}
\newcommand{\sref}[1]{Sec.~\ref{sec:#1}}
\newcommand{\tref}[1]{Table~\ref{tab:#1}}  

\def\TY#1{{\bf  \textcolor{red}{[TY: {#1}]}}}
\newcommand{\draftnote}[1]{{\bf\color{blue} #1}}
\newcommand{\draftnoteR}[1]{{\bf\color{red} #1}}

\title{ {\bf \Large{Light (and darkness) from a light hidden Higgs}\normalsize}}
\author{\bf{Roberto Vega$\,^{a}$,~Roberto Vega-Morales$\,^{b}$,~Keping Xie$\,^{a}$
}}

\vspace{-.1cm}
\affiliation{
$^a$Department of Physics, Southern Methodist University, Dallas, TX 75275, USA\\
$^{b}$Departamento de F\'{i}sica Te\'{o}rica y del Cosmos, Universidad de Granada,\\
Campus de Fuentenueva, E-18071 Granada, Spain}

\begin{abstract}
We examine light diphoton signals from extended Higgs sectors possessing (approximate) fermiophobia with Standard Model (SM) fermions as well as custodial symmetry.~This class of Higgs sectors can be realized in various beyond the SM scenarios and is able to evade many experimental limits, even at light masses, which are otherwise strongly constraining.~Below the $WW$ threshold, the most robust probes of the neutral component are di and multi-photon searches.~Utilizing the dominant Drell-Yan Higgs \emph{pair} production mechanism and combining it with updated LHC diphoton data, we derive robust upper bounds on the allowed branching ratio for masses between $45 - 160$~GeV.~Furthermore, masses $\lesssim 110$~GeV are ruled out if the coupling to photons is dominated by $W$ boson loops.~We then examine two simple ways to evade these bounds via cancellations between different loop contributions or by introducing decays into an invisible sector.~This also opens up the possibility of future LHC diphoton signals from a light hidden Higgs sector.~As explicit realizations, we consider the Georgi-Machacek (GM) and Supersymmetric GM (SGM) models which contain custodial (degenerate) Higgs bosons with suppressed couplings to SM fermions and, in the SGM model, a (neutralino) LSP.~We also breifly examine the recent $\sim 3\sigma$ CMS diphoton excess at $\sim 95$~GeV.
\end{abstract}

\preprint{UG-FT 327/18}
\preprint{CAFPE 197/18}
\preprint{SMU-HEP-18-08}
\preprint{FERMILAB-PUB-18-159-T}

\maketitle

\section{Introduction}\label{sec:Intro}

The nature of electroweak symmetry breaking (EWSB) appears to largely have been settled with the discovery of a 125~GeV scalar at the Large Hadron Collider (LHC)~\cite{Aad:2012tfa,Chatrchyan:2012xdj} possessing Standard Model (SM) Higgs boson like properties~\cite{Falkowski:2013dza}.~However, uncertainties in its coupling measurements~\cite{Khachatryan:2014kca,Khachatryan:2016vau,Sirunyan:2017exp,Sirunyan:2017tqd,Aaboud:2017oem,Blasi:2017xmc} still leaves room for extended Higgs sectors which can contribute non-negligibly to EWSB if they respect the well known `custodial' $SU(2)_C$ global symmetry~\cite{Sikivie:1980hm}, thus ensuring a tree level $\rho$ parameter equal to one.~These custodial Higgs bosons\,\footnote{We utilize the label Higgs boson for the neutral component which obtains a vacuum expectation value (VEV) as well as the charged components belonging to the same electroweak multiplet.} can have degenerate or compressed mass spectra, making them harder to detect due to soft decay products~\cite{Buckley:2009kv,Schwaller:2013baa,Ismail:2016zby,Egana-Ugrinovic:2018roi}.~Furthermore, as emphasized in~\cite{Delgado:2016arn}, if they have vanishing couplings to SM fermions (fermiophobic), and as measurements of the 125~GeV Higgs boson~\cite{Khachatryan:2014kca} are found to be more and more SM-like, previous searches which relied on single Higgs production mechanisms~\cite{Abreu:2001ib,Abbiendi:2002yc,Heister:2002ub,Achard:2002jh,Abbott:1998vv,Affolder:2001hx,Chatrchyan:2013sfs} become increasingly obsolete.~Thus, even for masses well below the 125~GeV Higgs boson, many limits which typically apply to extended Higgs sectors, can be evaded\,\footnote{Even a charged Higgs boson around or below the $W$ boson mass, which is not possible in the minimal supersymmetric model (MSSM)~\cite{Akeroyd:1995hg,Martin:1997ns}, is not excluded~\cite{Ilisie:2014hea,Enberg:2016ygw,Degrande:2017naf,Arbey:2017gmh,Arhrib:2017wmo}.}.

However, as shown in past~\cite{Barger:1992ty,Mrenna:2000qh,Landsberg:2000ht,Arhrib:2003ph,Akeroyd:2007yh,Akeroyd:2012ms} as well as more recent studies, diphoton~\cite{Delgado:2016arn,Degrande:2017naf} and multiphoton~\cite{Arhrib:2017wmo} searches can put robust constraints on these light, but otherwise difficult to detect Higgs bosons.~This is especially true when combined with the universal Drell-Yan Higgs \emph{pair} production mechanism~\cite{Aaltonen:2016fnw,Delgado:2016arn} which dominates for small exotic Higgs VEV.~Utilizing this, we combine Drell-Yan pair production with updated data from (inclusive) LHC diphoton searches~\cite{Aad:2014ioa,CMS-PAS-HIG-17-013} to derive robust upper bounds on the allowed branching ratio for masses between $45 - 160$~GeV.~We find the branching ratios must be $\lesssim 2 - 50\%$ depending on the mass and custodial representation.~Furthermore, if the coupling to photons are dominated by $W$ boson loops, custodial fermiophobic Higgs bosons are ruled out below $\sim 110$~GeV.

We then explore two simple ways to evade these bounds through cancellations between different loop contributions to the diphoton effective coupling and/or by introducing an invisible decay.~This also opens up the possibility of future LHC diphoton signals from a light exotic Higgs sector.~As part of our analysis we briefly explore the recently observed $\sim 3\sigma$ diphoton excess by CMS~\cite{CMS-PAS-HIG-17-013} at $\sim 95$~GeV, also examined in recent studies~\cite{Fox:2017uwr,Haisch:2017gql,Mariotti:2017vtv,Richard:2017kot,Liu:2018xsw}.~Finally, we examine two explicit realizations of these light Higgs sectors in the Georgi-Machacek (GM) and Supersymmetric GM (SGM) models~\cite{Vega:2017gkk} which contain custodial Higgs bosons with small couplings to SM fermions and, in the SGM model, an invisible (neutralino) LSP.
%


\section{Diphoton limits on custodial
fermiophobic Higgs bosons}\label{sec:light bosons}

After briefly reviewing custodial fermiophobic Higgs bosons, following closely the discussion in~\cite{Delgado:2016arn}, we then obtain limits from 8 and 13 TeV LHC \emph{inclusive} diphoton searches~\cite{Aad:2014ioa,CMS-PAS-HIG-17-013} on the allowed branching ratio into photons in the mass range $45-160$~GeV.~We also estimate what size branching ratios are needed to explain the recent $\sim 95$~GeV CMS diphoton excess~\cite{CMS-PAS-HIG-17-013}.

\subsection{Custodial fermiophobic Higgs sectors}\label{sec:prod}

Extended Higgs sectors that include only electroweak doublets with SM like quantum numbers automatically preserve custodial symmetry giving $\rho_{tree} = 1$, regardless of whether each doublet obtains the same VEV or not~\cite{Low:2010jp}.~However, since these can have tree level couplings to SM fermions, one is led to consider a `fermiophobic' limit to avoid constraints.~This limit is possible in certain Higgs doublet models such as the Type I two Higgs doublet model (2HDM)~\cite{Haber:1978jt,Mrenna:2000qh,Akeroyd:2010eg,Gabrielli:2012hd} or the `inert' 2HDM~\cite{Ma:2006km}, but \emph{not} the MSSM~\cite{Akeroyd:1995hg}.

To avoid resorting to highly tuned cancelations, larger electroweak representations are constrained by $\rho_{tree} = 1$ to come in ${\bf (N, \bar{N})}$ representations~\cite{Low:2010jp} of the global $SU(2)_L\otimes SU(2)_R$ symmetry (under which the SM Higgs is a ${\bf (2, \bar{2})}$) that breaks down to the custodial $SU(2)_C$ subgroup after EWSB.~The various Higgs bosons then decompose under the $SU(2)_C$ as $({\bf  N,\bar{N}}) = {\bf 1\oplus 3 \oplus 5  \oplus...}$ with the minimal case ${\bf N} = 3$ giving the GM model~\cite{Georgi:1985nv,Chanowitz:1985ug}, to be discussed more below.~In contrast to doublets, this requires multiple scalars for a given $SU(2)_L$ representation\,\footnote{For special representations satisfying the conditions derived in~\cite{Hally:2012pu,Low:2010jp}, such as an $SU(2)_L$ septet~\cite{Hally:2012pu,Alvarado:2014jva} with hypercharge $Y = 2$, this can be done with a single electroweak charged scalar.} with custodial symmetry ensuring their VEVs are `aligned' at tree level.~The various custodial scalars then exhibit (approximately) degenerate mass spectra between their neutral and charged components.

Since gauge invariance prevents a tree level coupling between these larger electroweak representations and SM fermions, any couplings to SM fermions are generated only by EWSB effects and suppressed by the exotic Higgs VEV and/or small mixing.~This leads to scalars which are naturally fermiophobic with respect to SM fermions.~These fermiophobic Higgs bosons have many generic phenomenological features which have been considered for some time~\cite{Pois:1993ay,Stange:1993ya,Diaz:1994pk,Akeroyd:1995hg,Akeroyd:1998dt,Akeroyd:1998ui,Barroso:1999bf,Brucher:1999tx,Landsberg:2000ht,Akeroyd:2003bt,Akeroyd:2003jp,Akeroyd:2003xi,Akeroyd:2005pr,Akeroyd:2010eg,Ilisie:2014hea} and searched for previously at LEP~\cite{Abreu:2001ib,Abbiendi:2002yc,Heister:2002ub,Achard:2002jh}, Tevatron~\cite{Abbott:1998vv,Affolder:2001hx}, and LHC~\cite{Chatrchyan:2013sfs}.~Since there is no coupling to SM fermions, there is no gluon fusion production available or corresponding decays.~Thus, large branching ratios into electroweak gauge bosons, in particular photons, are a generic feature if they are the lightest new particle~\cite{Delgado:2016arn}.~However, as we explore below, interference effects or if there is an exotic decay channel available, can dramatically alter this generic picture.

\subsection{Pair production and gauge boson decays}\label{sec:prod}

Any extension of the SM Higgs sector by electroweak charged scalars which contribute to EWSB will possess the Drell-Yan Higgs pair production channel $q\bar{q} \to W \to H_F^0 H_N^\pm$ which is \emph{not} present in the SM.~We take $H_F^0$ to generically represent a neutral fermiophobic Higgs boson while $H_N^{\pm}$ is in an arbitrary representation of the custodial $SU(2)_C$ symmetry labeled by $N$ which may or may not be in the same representation as $H_F^0$.~Although measurements of the 125~GeV Higgs boson couplings~\cite{Khachatryan:2014kca} still allow for non-negligible contributions to EWSB from fermiophobic Higgs sectors, already they constrain them enough that, at low masses, pair production dominates over single production channels which are suppressed by small VEVs~\cite{Delgado:2016arn}.~We write the $WHH$ vertex as,
\bea\label{eq:gwhh}
V_{WHH} \equiv i g 
\, C_N (p_1 - p_2)^\mu
\eea
where $C_N$ is determined by the $SU(2)_L$ representation and $p_{1}, p_{2}$ are the four momenta of the incoming and outgoing
scalar momenta.~When they are in different custodial representations, there is also a $Z$ mediated neutral Higgs pair production channel.

Since they are present in any custodial Higgs model with electroweak triplet representations or larger, we focus on the custodial singlet ($H_1$), triplet ($H_3$), and fiveplet ($H_5$) assuming they come from an electroweak bi-triplet $({\bf3,\bar{3}})$ which will also be examined in more detail below in the context of GM-type models.~The singlet and triplet could also appear in multi-Higgs doublet models\,\footnote{Of course they also appear in the SM where the Higgs boson decomposes as $({\bf2,\bar{2}}) = {\bf 1\oplus 3}$ under $SU(2)_C$, where the (approximate) custodial triplet gives the Goldstone bosons which become the longitudinal components of the $W$ and $Z$ bosons.}
with a fermiophobic limit~\cite{Akeroyd:1995hg}, though in this case the custodial (degenerate spectra) limit\,\footnote{Note that while the MSSM does not contain a fermiophobic limit~\cite{Akeroyd:1995hg}, it does have a custodial limit~\cite{Drees:1990dx} with $\tan\beta = 1$.}
is not necessary for $\rho_{tree} = 1$~\cite{Low:2010jp}.~However, CDF four photon searches~\cite{Aaltonen:2016fnw} more strongly constrain cases with a sizable mass splitting between the neutral and charged components.

In addition to the $WHH$ vertex in~\eref{gwhh}, $H_F^0$ will have couplings to $WW$ and $ZZ$ pairs which are generated during EWSB and which will be proportional to the \emph{exotic} Higgs vev~\cite{Georgi:1985nv,Akeroyd:2003bt,Akeroyd:2010eg,Cort:2013foa,Hartling:2014zca}.~We can parametrize these couplings generically with the following lagrangian,
\bea
\label{eq:LZW}
\mathcal{L}
&\supset&
s_\theta
\frac{H_F^0}{v} 
\Big( g_{Z} m_Z^2 Z^\mu Z_\mu + 2 g_{W} m_W^2 W^{\mu+} W^-_{\mu} 
\Big) ,
\eea
where $g_Z$ and $g_W$ are fixed by the $SU(2)_L \otimes U(1)_Y$ representation to which $H_F^0$ belongs.~The factor of $s_\theta$ simply parametrizes the `VEV mixing angle' or relative contribution to EWSB from the exotic Higgs VEV.~This ensures that as the exotic Higgs VEV tends to zero ($s_\theta \to 0$) the $H_F^0 VV$ couplings vanish along with all single production channels.~Note we also neglect Higgs mixing, which in the models we consider~\cite{Cort:2013foa} also goes to zero as $s_\theta \to 0$.~There may also be Higgs mixing generated during EWSB if there are multiple scalars in the same custodial representation or from custodial breaking effects at one loop, but these are neglected so that no Higgs mass mixing angles enter into~\eref{LZW}.~This also implies that any mixing with the SM-like 125~GeV Higgs boson is small as currently implied by Higgs couplings measurements~\cite{Khachatryan:2014kca}.~The ratio of the $g_Z$ and $g_W$ couplings,
\bea\label{eq:lamWZ}
\lambda_{WZ} = g_{W}/g_{Z},
\eea
is an important quantity~\cite{Chen:2016ofc} and is fixed by custodial symmetry at tree level to be $\lambda_{WZ} = 1$ or $\lambda_{WZ} = -1/2$ for a custodial singlet and fiveplet respectively~\cite{Low:2010jp}.~Note also that the factor of $s_\theta$ cancels explicitly in~\eref{lamWZ}.~While custodial triplets generically have vanishing tree level couplings~\cite{Cort:2013foa} to $WW$ and $ZZ$, the limits on diphoton branching ratios we obtain only depend on the pair production cross section so we include the triplet case in our analysis as well.~A more dedicated study of these `pseudo scalar' Higgs bosons would also be interesting.

At one loop the $g_W$ couplings in~\eref{LZW} will also generate effective couplings to $\gamma\gamma$ and $Z\gamma$ pairs (as well as $WW$ and $ZZ$) via $W$ boson loops.~We parametrize them with the dimension five effective operators,
\bea
\label{eq:LZA}
\mathcal{L}
&\supset&
\frac{H_F^0}{v} 
\Big( 
\frac{c_{\gamma\gamma} }{4} 
F^{\mu\nu} F_{\mu\nu} +
\frac{c_{Z\gamma}}{2} 
Z^{\mu\nu} F_{\mu\nu} 
\Big) ,
\eea
where $V_{\mu\nu}=\partial_\mu V_\nu - \partial_\nu V_\mu$ and we have assumed a CP even scalar.~Defining similar ratios,
\bea\label{eq:lamVA}
\lambda_{V\gamma} = c_{V\gamma}/\bar g_{Z},
\eea
where $V = Z, \gamma$ and we have implicitly absorbed a factor of $s_\theta$ into $\bar g_Z \equiv (s_\theta g_{Z})$.~There are also contributions to the effective couplings in~\eref{LZA} from additional charged Higgs bosons which are necessarily present, but typically subdominant to the $W$ vector boson loop.

\subsection{LHC diphoton limits and 95~GeV excess}\label{sec:LHC}

Surprisingly, the lone experimental search to utilize the Drell-Yan Higgs pair production channel and combine it with (multi)photon searches for a light fermiophobic Higgs boson is a recent CDF analysis of previously collected Tevatron $4\gamma + X$ data~\cite{Aaltonen:2016fnw}.~However, this search relies on the decay of the charged Higgs boson to the neutral Higgs being kinematically available.~Thus, in the limit where the mass splitting between the pair of Higgs bosons goes to zero, limits from this multiphoton search can be evaded.~In models with custodial symmetry~\cite{Sikivie:1980hm} in the Higgs sector, which are motivated by electroweak precision data, degenerate masses between the neutral and charged Higgs bosons are generated (at tree level).~This makes the CDF four photon search insensitive to \emph{custodial} fermiophobic Higgs bosons~\footnote{Of course if there are additional Higgs scalars which are in different custodial representations than $H_F^0$, additional Higgs pair production mechanisms with non-degenerate masses can become available allowing for $4\gamma + X$ limits to again be applied.}.~As emphasized in~\cite{Delgado:2016arn}, diphoton searches have the advantage that, being more inclusive, are more model independent and can be applied even in the custodial limit of degenerate masses as well as when $M_{H^{\pm}} < M_{H^{0}}$ or if the charged Higgs decays in a way that is difficult to observe.

Combining updated 8 and 13~TeV low mass diphoton data~\cite{Aad:2014ioa,CMS-PAS-HIG-17-013}, we can obtain new robust bounds on the allowed branching ratio into photons for different cases of custodial fermiophobic Higgs bosons in the mass range $65 - 160$~GeV.~For the necessary production channels we have used a modified version of the Madgraph~\cite{Alwall:2014hca} framework developed for the GM model in~\cite{Hartling:2014xma} to compute cross sections at leading order for an 8 and 13~TeV LHC.~There are $\mathcal{O}(1)$ largely model independent $k$-factors~\cite{Ilisie:2014hea,Degrande:2015xnm} arising from corrections which are neglected, but this will not qualitatively change our results and can easily be included in a more precise analysis.

\begin{figure}[tbh]
\begin{center}
\includegraphics[scale=.49]{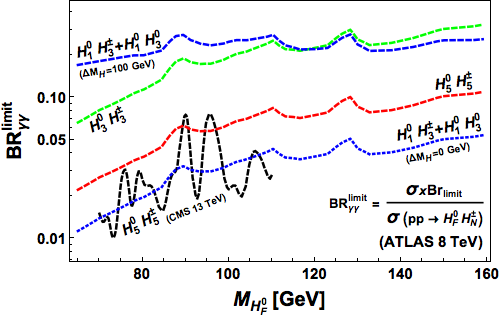}
\end{center}
\caption{The dashed colored lines show the allowed branching ratio by 8~TeV ATLAS diphoton searches~\cite{Aad:2014ioa} with $20.3\,fb^{-1}$ of data ($65 - 160$~GeV) as a function of mass for a custodial fermiophobic Higgs boson produced dominantly via the Drell-Yan $pp\to W^{\pm} \to H_F^0 H_N^{\pm}$ Higgs pair production channel.~The custodial singlet ($H_1^0$), triplet ($H_3^0$), and fiveplet ($H_5^0$) cases are shown with couplings defined in~\eref{gwhh}.~For the range $70-110$~GeV, we also show (black dashed) the more recent $13$~TeV CMS low mass diphoton search~\cite{CMS-PAS-HIG-17-013} which has a $\sim 3\sigma$ excess at $\sim95$~GeV with $35.9\,fb^{-1}$ of data.}
\label{fig:HBR}
\end{figure}
We show in~\fref{HBR} the allowed branching ratio (dashed colored lines) by 8~TeV ATLAS (inclusive) diphoton searches~\cite{Aad:2014ioa} in the range $65 - 160$~GeV.~For the fiveplet in the range $70-110$~GeV, we also show (black dashed) the more recent $13$~TeV CMS low mass diphoton search~\cite{CMS-PAS-HIG-17-013} which has a $\sim 3\sigma$ excess at $\sim95$~GeV with $35.9\,fb^{-1}$ of data.~We see that for the fiveplet, with group theory factor $C_5 = \sqrt{3}/2$, branching ratios $\gtrsim 2-3\%$ are excluded in the region below the $W$ mass.~At masses above $150$~GeV, they can be large as $\sim 10\%$.~For a custodial triplet, bounds are a bit weaker due to the smaller group theory factor $C_3 = 1/2$.~In this case branching ratios up to $\sim 5-10\%$ are still allowed at low masses while at high masses they can be as large as $\sim 30\%$.

For the custodial singlet (blue), custodial symmetry restricts the singlet to be pair produced with a custodial triplet and gives $C_1 = \sqrt{2/3}$.~In this case a $Z$ boson mediated channel also opens up which has been included.~Generically the triplet has a different mass than the singlet.~For this we consider two cases; one where the singlet and triplet are degenerate (dotted) and one where we take the triplet to be $100$~GeV heavier (dashed).~Due to the additional production channel, we see for the degenerate case better sensitivity than for the fiveplet, with branching ratios greater than $\sim 1-2\%$ ruled out in the low mass region.~When there is a $100$~GeV splitting, branching ratios as large as $\sim 15 - 20\%$ are allowed for the custodial singlet at low masses and furthermore, the weak dependence on the $H_1^0$ mass.~Note this size of mass splitting is just at the edge of the largest splitting which can be probed by the CDF multiphoton search~\cite{Aaltonen:2016fnw}.

We also see in~\fref{HBR} the need to extend 13 TeV diphoton searches to cover the window between $110$~GeV and the lower cutoff of $200$~GeV for higher mass searches at 13~TeV~\cite{Aaboud:2016tru}.~As emphasized in~\cite{Mariotti:2017vtv}, extending and optimizing diphoton searches below $65$~GeV could also be greatly beneficial as neutral (and charged) Higgs bosons which may have escaped detection, perhaps all the way down to half of the $Z$ (and $W$) mass, are in principle still possible~\cite{Ilisie:2014hea,Enberg:2016ygw,Degrande:2017naf,Arbey:2017gmh,Arhrib:2017wmo}.~Note that bounds for the custodial singlet and triplet can be mapped onto 2HDMs with appropriate rescaling by mixing angles~\cite{Akeroyd:2003bt,followup}.

From~\fref{HBR} we can also assess roughly what size branching ratios are needed to explain the $\sim 3 \sigma$ diphoton excess at $\sim 95$~GeV recently observed by CMS~\cite{CMS-PAS-HIG-17-013} and corresponding to a cross section of $\mathcal{O}(0.05-0.1)$\,pb~\cite{Fox:2017uwr}.~Assuming production is dominated by the Drell-Yan mechanism discussed above, this implies that if the excess is due to a custodial fiveplet Higgs boson, $\sim 5\%$ diphoton branching ratios are needed.~For the custodial triplet we find (but do not plot) branching ratios around $\sim 20\%$ are needed.~For the two singlet cases, degenerate and $\Delta M_H = 100$~GeV, one needs branching ratios around $\sim 3\%$ and $30\%$ respectively.~How easily these branching ratios can be achieved depends on a particular model, but are generically achievable for fermiophobic Higgs bosons unlike those with SM-like Higgs boson branching ratios~\cite{Barger:1992ty} which are far too small.~Once backgrounds are accounted for, the branching ratio needed is likely smaller, but a rough estimate based on a conservative upper bound is sufficient for present purposes.

In the low mass region of the diphoton search window considered here, various limits on charged Higgs bosons from LEP in principle apply, but these can be evaded if the charged Higgs is fermiophobic~\cite{Arhrib:2006wd,Enberg:2013jba,Ilisie:2014hea,Degrande:2017naf}.~The same is true for indirect constraints such as $b\to s\gamma$~\cite{Hartling:2014aga}.~For a custodial fiveplet, same sign dilepton searches for a doubly charged scalar rule out masses below $\sim 76$~GeV assuming $100\%$ branching ratio into same sign $W$ bosons~\cite{Logan:2015xpa} which may or may not hold in specific models~\cite{Cort:2013foa,Garcia-Pepin:2014yfa,Vega:2017gkk}.~Contributions to exotic decays of the SM-like Higgs boson~\cite{Curtin:2013fra} and $Z$ boson~\cite{Blinov:2017dtk} will be relevant for light enough masses and deserves further investigation.

Given current constraints on the 125~GeV Higgs boson couplings~\cite{Khachatryan:2014kca}, the bounds obtained in~\fref{HBR} are already stronger than those obtained assuming SM-like production mechanisms~\cite{Abreu:2001ib,Abbiendi:2002yc,Heister:2002ub,Achard:2002jh,Abbott:1998vv,Affolder:2001hx,Chatrchyan:2013sfs} and will get increasingly so as time goes on without observing a deviation from SM-like Higgs boson couplings.~These diphoton searches can be replaced by, or combined with, inclusive searches in other final states as well as be combined with searches for the charged components.~Because of the universal nature of the Drell-Yan pair production channel, this allows for the possibility of a powerful and model independent probe of extended Higgs sectors.~Furthermore, the much larger production cross sections at future colliders~\cite{Dawson:2013bba} would allow for an especially powerful probe of these potentially hidden exotic Higgs sectors.~We leave an exploration of these interesting possibilities to ongoing work~\cite{followup}.

\subsection{Escaping current $\&$ future diphoton limits}\label{sec:limits}

As discussed, two simple ways to evade these bounds are via cancellations between different loop contributions to the diphoton decay~\cite{Akeroyd:2007yh} or by introducing an invisible decay into a dark sector.~Focusing first on the former we again show in~\fref{HBRlam} the allowed branching ratio into photons as a function of mass in the range $45 - 160$~GeV.~The top set of dashed colored lines are the same as in~\fref{HBR} while the lower colored dashed lines are the same as the top ones but (naively) projected (neglecting CMS search) assuming a two orders of magnitude improvement in sensitivity.~While this sensitivity is beyond the future reach of LHC diphoton searches~\cite{Mariotti:2017vtv,Dawson:2013bba}, it should be achievable at future high energy colliders~\cite{Dawson:2013bba}.
\begin{figure}[tbh]
\begin{center}
\includegraphics[scale=.49]{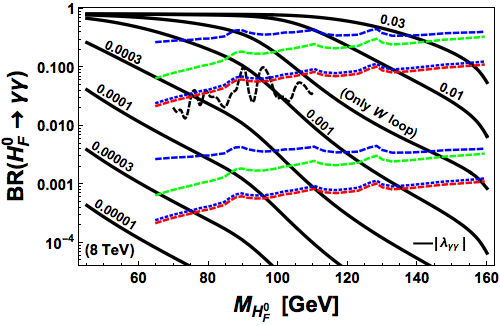}
\end{center}
\caption{The top set of dashed colored lines are the same as in~\fref{HBR}.~The lower colored dashed lines (neglecting CMS search) are the same as the top, but projected assuming a two orders of magnitude improvement in sensitivity.~The black solid lines indicate contours of the effective coupling ratio $\lambda_{\gamma\gamma}$ defined in~\eref{lamVA}.~We also indicate the contour corresponding to only the $W$ loop contribution to the effective coupling.}
\label{fig:HBRlam}
\end{figure}

The black solid lines indicate contours of the effective diphoton coupling ratio $\lambda_{\gamma\gamma}$ defined in~\eref{lamVA} (we have also set $\lambda_{\gamma\gamma} = \lambda_{Z\gamma}$).~As in~\cite{Delgado:2016arn}, to compute the diphoton branching ratio for these contours, we have included decays to $\gamma\gamma$, $Z\gamma$, $WW$, and $ZZ$ in the total decay width for a neutral fermiophobic Higgs boson.~To obtain the necessary three and four body decays we have integrated the analytic expressions for the $H_F^0 \to V\gamma \to 2\ell\gamma$ and $H_F^0 \to VV \to 4\ell$ fully differential decay widths computed and validated in~\cite{Chen:2012jy,Chen:2013ejz,Chen:2014ona}.~For the explicit $W$ loop functions which contribute to the effective couplings we use the parametrization and implementation found in~\cite{Chen:2015rha}.~We also indicate the contour corresponding to only the $W$ loop contribution to the effective coupling assuming a custodial fiveplet with $\lambda_{WZ} = -1/2$.~We find similar contours for the case of a singlet with $\lambda_{WZ} = 1$.

We see that, as found in~\cite{Delgado:2016arn}, if the $W$ boson loop dominates the effective coupling to photons,~8~TeV LHC diphoton searches~\cite{Aad:2014ioa} rule out a custodial fermiophobic Higgs boson below $\sim 110$~GeV.~This is just at the upper limit of the more recent $13$~TeV CMS low mass diphoton searches~\cite{CMS-PAS-HIG-17-013} in the range $70-110$~GeV.~We also see that if a future collider is able to improve on current limits by two orders of magnitude, masses up to $\sim 150$~GeV can be ruled out in this scenario.~Above these masses, $ZZ$ and $WW$ searches typically become more powerful due to the diphoton branching ratio becoming too suppressed even for an enhanced effective coupling~\cite{Delgado:2016arn}.~Thus with a future collider, light custodial fermiophobic Higgs bosons can perhaps be completely ruled out below the diboson thresholds if their couplings to photons are dominated by $W$ boson loops.~We also emphasize that in this case, the limits are independent of the exotic Higgs VEV as it cancels explicitly in any of the branching ratios~\cite{Delgado:2016arn}. 

As can also be seen, for values of the effective coupling ratio $\lambda_{\gamma\gamma} \lesssim 10^{-3}$ one can lower the limit from LHC diphoton searches, perhaps even below the lowest end of the current search window of 65~GeV when $\lambda_{\gamma\gamma} \lesssim 10^{-4}$.~These suppressions require $\mathcal{O}(1-50\%)$ level cancelations between the $W$ boson loop and other contributions.~Though this implies a certain level of tuning, it can happen and in particular in models containing doubly charged particles~\cite{Degrande:2017naf}.~This also illustrates the importance of extending diphoton searches to as low a mass as possible~\cite{Mariotti:2017vtv} since masses below 65~GeV are not ruled out by Tevatron four photon searches~\cite{Aaltonen:2016fnw} in the degenerate (custodial) limit.~We also see that $|\lambda_{\gamma\gamma}| \sim 0.001$ is needed to explain the 95~GeV excess which requires $\sim 40\%$ level cancelations.~When there is large destructive interference between the different loop contributions to the diphoton effective coupling, (off-shell) $ZZ$ and $WW$ three and four body decays can be sizable in the low mass region.~Thus also extending $ZZ$ and $WW$ searches~\cite{Chatrchyan:2013sfs} as low as possible is crucial for closing any allowed windows.

Larger values of $\lambda_{V\gamma} \gtrsim 0.01 $ are also possible allowing for larger masses to be excluded.~Such large values for this ratio can easily be obtained~\cite{Delgado:2016arn} in the limit $s_\theta \ll 1$ if there exist additional mass scales apart from the Higgs VEVs in the scalar potential or if the loop particles carry large charges.~In this case of enhanced couplings to photons, the diphoton channel can be sizable all the way up to the $WW$ threshold~\cite{Delgado:2016arn}.~The $H_F^0 \to V\gamma \to 2f\gamma$ three body decay through an off-shell photon or $Z$ can also be sizable up to $\sim 130$~GeV and should be studied further.

In addition to loop cancelations, a second and more natural way of evading these constraints is to allow for the possibility of an exotic, and in particular, invisible decay which suppresses the branching ratio to photons.~Below we explore two explicit realizations of these possibilities for evading diphoton constraints in the GM and SGM models which contain custodial fermiophobic Higgs bosons and, in the case of the SGM, an invisible LSP.


\section{Light Signals in the GM and\\ Supersymmetric GM Model}\label{sec:sgm}

The GM model~\cite{Georgi:1985nv,Chanowitz:1985ug} is one of the most thoroughly explored examples of an extended (non-doublet) Higgs sector containing custodial fermiophobic Higgs bosons.~This model has been shown to have a rich phenomenology~\cite{Chanowitz:1985ug,Gunion:1989ci,Gunion:1990dt} which has been examined in many recent studies~\cite{Englert:2013zpa,Chiang:2012cn,Chiang:2014bia,Chiang:2015kka,Chiang:2015amq,Degrande:2015xnm,Hartling:2014zca,Hartling:2014aga,Degrande:2015xnm,Logan:2015xpa,Campbell:2016zbp,deFlorian:2016spz,Degrande:2017naf,Logan:2017jpr,Zhang:2017och}.~In minimal versions~\cite{Gunion:1989ci,Hartling:2014aga}, there is no neutral LSP which could open up an invisible decay channel to avoid diphoton constraints.~In this case, cancellations are needed to suppress the diphoton branching ratio sufficiently.~However, the presence of doubly charged scalars in the model allows for larger destructive interference with $W$ boson loops which can lead to a suppressed effective coupling to photons.~These cancellations have also been shown~\cite{Degrande:2017naf} to open up the possibility of avoiding stringent LEP diphoton search constraints (when $s_\theta \gtrsim 0.1$) for masses below $\sim 110$~GeV.~One could also simply add an additional stable neutral particle giving a potential dark matter candidate~\cite{Campbell:2016zbp} and opening up an invisible decay channel.

Supersymmetric models naturally give ways to have an extended Higgs sector with an invisible sector to decay into.~However since there is no fermiophobic limit in the MSSM~\cite{Akeroyd:1995hg}, a light diphoton signal, such as the 95~GeV CMS excess~\cite{CMS-PAS-HIG-17-013}, is likely difficult to reconcile, but can perhaps be explained in Type-1 2HDM models~\cite{Fox:2017uwr,Haisch:2017gql} (or the 'natural' NMSSM~\cite{Cao:2016uwt}).~Thus one is led to consider extended MSSM Higgs sectors.~Extensions of the MSSM Higgs sectors have of course been considered many times to alleviate difficulties in the MSSM with explaining the observed 125~GeV Higgs boson mass without resorting to heavy stops~\cite{Draper:2011aa}.~We consider one such case in the Supersymmetric Custodial Higgs Triplet Model (SCTM)~\cite{Cort:2013foa,Garcia-Pepin:2014yfa,Delgado:2015bwa}, constructed to alleviate the MSSM Higgs mass `problem' while at the same time satisfying constraints from EWPD and other direct searches.

As shown in~\cite{Vega:2017gkk}, the SCTM has a low energy limit, which defines the SGM, that gives rise to the same Higgs boson sector as in the GM model, but also includes the presence of light fermionic superpartners.~The SGM also inherits all of the other attractive features of the SCTM~\cite{Cort:2013foa,Garcia-Pepin:2014yfa,Delgado:2012sm,Delgado:2015aha,Delgado:2015bwa,Carena:2013ooa,Carena:2014nza,Carena:2015moc,Garcia-Pepin:2016hvs}.~In the SGM model there is of course the possibility of cancellations between $W$ boson loops and doubly charged scalars, but now also with doubly charged fermions.~The neutralino sector provides an invisible sector for the scalar Higgs bosons to potentially decay into and in particular, a light (neutralino) LSP.~To explore this we perform various scans to find regions of parameter space which can escape LHC diphoton searches in the $45-160$~GeV mass range.~We also briefly examine the recently observed 95~GeV CMS diphoton excess~\cite{CMS-PAS-HIG-17-013}.~First we breifly review the GM and SGM models, but refer the reader to~\cite{Hartling:2014aga,Vega:2017gkk} for details.

\subsection{Lightning review of GM $\&$ SGM models}\label{sec:sgm}

In the minimal GM model, on top of the SM Higgs doublet $H=(H^+,H^0)^T$, one real $SU(2)_L$ triplet scalar with hypercharge $Y=0$, $\phi=(\phi^+,\phi^0,\phi^-)^T$, and one complex triplet scalar with $Y = 1$,  $\chi=(\chi^{++},\chi^+,\chi^0)^T$, are added.~In terms of representations of $SU(2)_L\otimes SU(2)_R$ we have the $2\times 2$ and $3\times 3$ matrix fields,
\begin{equation}
\label{eq:HandX}
H=\left(\begin{matrix}
H^{0*} & H^+\\
H^-&H^0
\end{matrix}  
\right),\quad
X=\left(\begin{matrix}
\chi^{0*} & \phi^+& \chi^{++}\\
\chi^-&\phi^0&\chi^+\\
\chi^{--}&\phi^-&\chi^0 
\end{matrix}  
\right),
\end{equation}
transforming as $({\bf 2,\bar2})$ and $({\bf 3,\bar3})$, respectively.~If EWSB proceeds such that $v_H\equiv \langle H^0\rangle$, $v_X\equiv\langle \phi^0\rangle=\langle \chi^0\rangle$,~i.e.~the triplet VEVs are aligned, then the $SU(2)_L\otimes SU(2)_R$ will be broken to the custodial subgroup $SU(2)_C$, which ensures that the $\rho_{tree} = 1$ as in the SM~\cite{Low:2010jp}.~The bi-doublet and bi-triplet Higgs fields then decompose under the $SU(2)_C$ as $({\bf2,\bar{2}}) = {\bf 1\oplus 3}$ and $({\bf3,\bar{3}}) = {\bf 1\oplus 3 \oplus 5}$.~This global symmetry breaking structure can also be imbedded into certain composite Higgs models~\cite{Georgi:1985nv,Chanowitz:1985ug,Bellazzini:2014yua}.

Using similar conventions to~\cite{Hartling:2014zca}, we can write the $SU(2)_L\otimes SU(2)_R$ invariant GM model Higgs potential,
\bea
\label{eq:vgm} 
V_{GM} &=& 
\frac{\mu_2^2}{2} \Tr{ {H}^\dagger {H} } 
+ \frac{\mu_3^2}{2} \Tr{ {X}^\dagger {X} }\nn
&+&  \lambda_1 \Tr{ {H}^\dagger {H}  }^2 
+ \lambda_2 \Tr{ {H}^\dagger {H} }\, \Tr{ {X}^\dagger {X} } \nn
&+& \lambda_3 \Tr{ {X}^\dagger X {X}^\dagger {X} } 
+ \lambda_4  \Tr{ {X}^\dagger {X} }^2 \\
&-& \lambda_5 \Tr{ {H}^\dagger \tau^a  {H}  \tau^b } 
\Tr{ {X}^\dagger t^a {X} t^b  } \nn
&-& M_1 \Tr{ {H}^\dagger \tau^a {H}  \tau^b } (U {{X}} U^\dagger)_{ab} \nn
&-&  M_2 \Tr{ {X}^\dagger t^a  {X} t^b } (U {\bar{X}} U^\dagger)_{ab} , \nonumber
\eea
where $\tau_i = \sigma_i/2$ and $t_i$ are the two and three dimensional representations respectively of the $SU(2)$ generators.~As shown in~\cite{Vega:2017gkk} and discussed above, the potential in~\eref{vgm} can be `derived' from the Higgs potential of the SCTM~\cite{Cort:2013foa,Garcia-Pepin:2014yfa,Delgado:2015bwa}.~However, its supersymmetric origin leads to the constraints on the quartic couplings~\cite{Vega:2017gkk},
\bea
\label{eq:gmcon}
\lambda_1 &=& \frac{3}{4} \lambda_2,\,\lambda_3 = -\lambda_4,\\
\lambda_5 &=& -4\lambda_2 + 2\sqrt{2\lambda_2\lambda_4} , \nonumber
\eea
reducing the number of quartics from five to two.

Once the electroweak symmetry breaking conditions~\cite{Cort:2013foa,Hartling:2014zca} and constraints in~\eref{gmcon} are enforced, we have six free Higgs potential parameters given by,
\bea\label{eq:params}
(\lambda_2,\,\lambda_4,\,M_1,\,M_2,\,v_H,\,v_X).
\eea
When the trilinear soft breaking mass parameters are small in the SCTM, such as in the gauge mediated symmetry breaking scenario~\cite{Delgado:2015bwa}, there is a \emph{one-to-one} correspondence between the six free parameters in~\eref{params} and the four superpotential parameters plus Higgs doublet and triplet VEV's in the SCTM~\cite{Vega:2017gkk}.~Thus the SGM can be seen as a weak scale effective theory given by the GM model, with the constraint in~\eref{gmcon} applied, plus custodial fermions at the same scale as the custodial Higgs bosons.~As examined in~\cite{Vega:2017gkk}, in the `slice' of parameter space defined by~\eref{gmcon}, the GM model can appear to be very similar to the SGM model depending on the exact masses of the fermion superpartners.~As in~\cite{Vega:2017gkk}, we consider the constrained GM model when comparing to the LHC phenomenology of the SGM model.

We can also use the constraint from EWSB on the doublet and triplet VEVs which requires them to satisfy~\cite{Vega:2017gkk},
\bea\label{eq:vevdef}
v^2 &=& 2v_H^2 + 8 v_X^2 = \frac{4 m_W^2}{g^2} ,
\eea
and leads to an explicit definition for the mixing angle defined in~\eref{LZW}, $s_\theta \equiv 2\sqrt{2}\,v_X/v$.~Then, using measurements~\cite{Agashe:2014kda} of the Higgs and $W$ boson masses as well as electroweak gauge couplings to impose $v = 246\, {\rm GeV},~m_{h} = 125 \, \mbox{GeV}$, we can eliminate two parameters in~\eref{params}.~Below we perform various scans in the resulting four dimensional parameter space.

In the SGM there is of course the presence of the gaugino/higgsino sector coming from the SCTM~\cite{Cort:2013foa} which can also be examined in terms of custodial symmetry~\cite{Vega:2017gkk}.~Thus like the scalar Higgs bosons, the higgsinos can be arranged into a custodial singlet and triplet coming from the (MSSM) electroweak doublets and a custodial singlet, triplet, and fiveplet coming from the electroweak triplets.~Furthermore, the Higgsino masses are determined by the Higgs potential parameters in~\eref{params} and thus correlated with the Higgs scalar masses.~There are also the gauginos which we take to be much heavier than the weak scale higgsinos as in~\cite{Vega:2017gkk}.~Over some regions of parameter space, the lightest neutralino can make a viable thermal dark matter candidate~\cite{Delgado:2015aha}.

In general these fermions can be produced in pairs via Drell-Yan, but can be difficult to detect due to their compressed spectra~\cite{Schwaller:2013baa,Ismail:2016zby} so are only constrained to be $\gtrsim 100$~GeV and perhaps even as low as $\sim 75$~GeV~\cite{Egana-Ugrinovic:2018roi}.~However, if the custodial fiveplet is the LSP constraints may be stronger~\cite{Ostdiek:2015aga}.~We do not conduct an in depth study of the gaugino/higgsino sector here since our focus is exploring its effects on the diphoton branching ratio of the lightest custodial Higgs boson.~A more in depth study examining LHC searches for gaugino/higgsinos with compressed spectra and combining them with other experimental constraints on the SGM model is ongoing~\cite{followup}.

\subsection{Fiveplet diphoton signals at the LHC}\label{sec:LHC}

In principle any of the (neutral) custodial scalars in the GM/SGM model can give a light diphoton signal.~However, as discussed, the custodial singlets and triplets coming from the electroweak doublet and triplets can mix~\cite{Cort:2013foa}.~This induces couplings to SM fermions, though they are suppressed by EWSB.~On the other hand, for the fiveplet, custodial symmetry prevents the neutral component ($H_5^0$) from mixing with other neutral scalars and in particular with the 125~GeV SM Higgs boson.~This allows for the fermiophobic condition to be maintained without fine tuning~\cite{Georgi:1985nv,Cort:2013foa,Hartling:2014zca} or resorting to renormalization conditions (as needed in two Higgs doublet models~\cite{Akeroyd:2010eg}).~Thus the custodial fiveplet in GM-type models is a naturally fermiophobic scalar\,\footnote{The physical $T$-odd scalar in Littlest Higgs Models with $T$-parity~\cite{Cheng:2003ju,Cheng:2004yc,Low:2004xc,Hubisz:2004ft}, which has zero VEV, resembles the custodial fiveplet with degenerate neutral and charged components.~However in this case, $T$-parity prevents decay to pairs of photons.} which can give rise to light diphoton signals at the LHC.

To explore this we perform various scans over the four dimensional parameter space in~\eref{params} after imposing measurements of the SM-like Higgs boson and the electroweak scale VEV, limiting us to $v_X \leq 15$~GeV $(s_\theta \lesssim 0.1)$.~This is still significantly larger than that allowed by electroweak precision data~\cite{Agashe:2014kda} for \emph{non}-custodial electroweak triplets whose VEV is restricted to $s_\theta  \lesssim 0.001$~\cite{Beringer:1900zz,Delgado:2012sm,Delgado:2013zfa,Garcia-Pepin:2014yfa}.~Similarly to~\cite{Vega:2017gkk}, we trade in one Higgs potential parameter to scan over the custodial fiveplet mass, while demanding perturbative quartic couplings~\cite{Hartling:2014zca} and mass parameters around the weak scale.~We assume the fiveplet is the lightest custodial Higgs boson which leads to $m_1, m_3 \gtrsim 130$~GeV for the singlet and triplet masses.~For the small $s_\theta$ range in which we work, bounds from direct and indirect constraints are easily evaded for this mass range~\cite{Hartling:2014aga,Degrande:2017naf}.~We limit ourselves to a leading order (custodial) analysis, but loop corrections to custodial Higgs boson masses can be large, and sometimes divergent, for heavy masses and large triplet VEVs in (non-supersymmetric) GM type models~\cite{Braathen:2017izn,Krauss:2017xpj}.~For all of the calculations needed to conduct our parameter scans we have used the SARAH/SPheno~\cite{Porod:2003um,Porod:2011nf,Staub:2015kfa} package and validated for a few random points with FeynArts/FormCalc/LoopTools~\cite{Hahn:1998yk,Klasen:2002xi}.

In our first scan we impose the additional constraint $|\lambda_2| = |\lambda_4|, |M_1| = |M_2|$ to conduct a finer two dimensional scan with $\Delta m_5 = 2$~GeV in the range $45  \leq m_5 \leq  160$~GeV.~Since it is more computationally intensive, we also conduct a less fine four dimensional scan in the range $50  \leq m_5 \leq 160$~GeV with $\Delta m_5 = 10$~GeV.~To explore the 95 GeV CMS diphoton excess, we also perform a four dimensional scan between $92 \leq m_5 \leq 98$~GeV with $\Delta m_5 = 2$~GeV.~The results from all three scans are combined into one and shown in~\fref{BrH5aaconts}.
\begin{figure}[tbh]
\includegraphics[width=.495\textwidth]{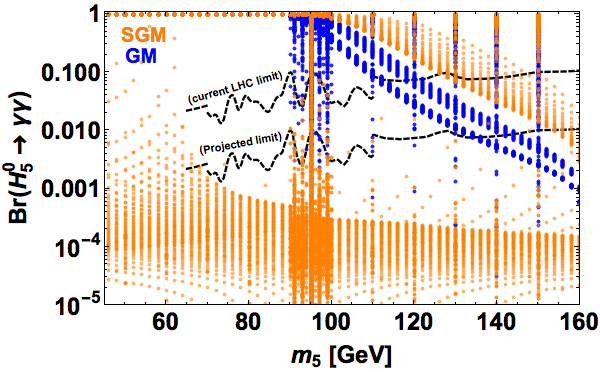}
\includegraphics[width=.48\textwidth]{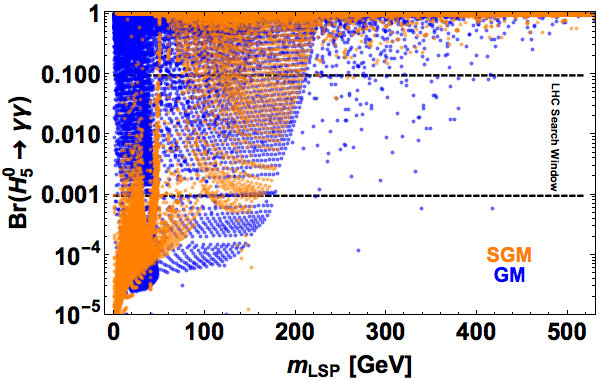}
\caption{{\bf Top:~}Custodial fiveplet branching ratio into photons in the GM model (blue) and SGM model (orange) as a function of fiveplet mass.~The current bounds from LHC diphoton data~\cite{Aad:2014ioa,CMS-PAS-HIG-17-013} are shown (top dashed curve) as well as a rough future projection of sensitivity (lower dashed curve) assuming an order of magnitude improvement at a high luminosity LHC~\cite{Dawson:2013bba}.~{\bf Bottom:~}Same as top, but as a function of the LSP mass.~The black dashed lines indicate a rough estimate of the potential LHC diphoton search `window'.}
\label{fig:BrH5aaconts}
\end{figure}

On top we show the fiveplet branching ratio into photons in the (constrained) GM model (blue) and SGM model (orange) as a function of the custodial fiveplet mass.~The current bounds are shown (top dashed curve) from $20\, fb^{-1}$ of 8 TeV ATLAS diphoton data~\cite{Aad:2014ioa} between $65-70$~GeV and $110-160$~GeV, combined with $35.9\,fb^{-1}$ of 13 TeV CMS diphoton data~\cite{CMS-PAS-HIG-17-013} from $70-110$~GeV.~To gain can an idea of future possibilities, we also show a rough future projection of sensitivity (lower dashed curve) assuming an order of magnitude improvement at a high luminosity LHC~\cite{Dawson:2013bba,Mariotti:2017vtv}.~On bottom we show the same, but as a function of the LSP mass from $2-520$~GeV and (roughly) indicate the potential LHC `window' of sensitivity.~For the GM model which does not have an LSP, the points correspond to the same value of Higgs potential parameters (see~\eref{vgm}) as in the SGM model, which in turn determines the (higgsino) LSP mass.~Thus the differences in parameter space are due to the effects from the higgsino sector, both via loop effects and, when light enough, opening up new decays.

The first thing to note is the power of diphoton searches to rule out much of the parameter space in both models when $Br(H_5\to\gamma\gamma)$ is $\mathcal{O}(1)$ which, as discussed above, is a generic feature of fermiophobic Higgs bosons~\cite{Delgado:2016arn}.~We also see the significantly larger parameter space in the SGM that is allowed by diphoton searches than for the GM mode.~This is due almost entirely to decays into the light LSP opening up since, in the SGM, the doubly charged scalar and higgsino fiveplets necessarily interfere destructively in the diphoton loops.~Thus, cancelation effects with the $W$ boson loops are generically smaller than even in the constrained GM model defined by~\eref{gmcon}.~We see this in the bottom of~\fref{BrH5aaconts} with the smaller allowed parameter space in the SGM model at larger LSP masses where suppression of the diphoton branching ratio becomes dominated by interference effects.~Of course, in the general GM model~\cite{Hartling:2014zca} even more parameter space should be available.

We also see (top) that a future high luminosity LHC may be able to rule out much of the currently allowed parameter space below $\sim160$~GeV after which $WW$ and $ZZ$ searches typically become more sensitive~\cite{Delgado:2016arn}.~In the SGM model, we see the presence of a light neutralino allows for very suppressed branching ratios, potentially evading even future LHC diphoton limits for branching ratios $\lesssim 10^{-4}$.~In this case, the lightest neutralino must be a custodial singlet due to constraints on light charged fermions~\cite{Schwaller:2013baa,Ismail:2016zby,Egana-Ugrinovic:2018roi}.~Missing energy searches for light dark matter~\cite{Goodman:2010yf} then become relevant and a dedicated study of these interesting possibilities is ongoing~\cite{followup2}.~A future high energy collider should probe and possibly rule out much of the remaining allowed parameter space.

Finally, for the 95~GeV CMS diphoton excess~\cite{CMS-PAS-HIG-17-013} we see (top) with our dense scan between $92 \leq m_5 \leq 98$~GeV that in both models there are parameter points which can accommodate the excess.~At this fiveplet mass, interference effects in both models dominates the suppression effect when $m_{LSP} \gtrsim 50$~GeV, at which point invisible two body decays are no longer available in the SGM.~In the SGM we also see that just at threshold as two body decays open up, the diphoton branching ratio is suppressed enough to not be ruled out, but still large enough to explain the excess.~In this case a $\sim 95$~GeV diphoton signal would imply a neutralino around $45 - 50$~GeV which could be targeted in LHC invisible searches~\cite{Goodman:2010yf,Rajaraman:2011wf}.~Once the LSP mass is lighter than this threshold, the branching ratio quickly becomes highly suppressed as seen in the threshold behavior around $50$~GeV in (bottom)~\fref{BrH5aaconts}.

\section{Conclusions}\label{sec:conc}

We have examined potential light diphoton signals at the LHC coming from custodial fermiophobic Higgs bosons in the mass range $45 - 160$~GeV.~We have emphasized that due to their lack of coupling to SM fermions and degenerate mass spectra, they can evade many of the stringent constraints which typically apply to extended Higgs sectors.~However, when combined with the dominant Drell-Yan Higgs pair production mechanism, diphoton searches at the LHC can provide robust constraints.~We have utilized this with 8 and 13~TeV LHC inclusive diphoton searches~\cite{Aad:2014ioa,CMS-PAS-HIG-17-013} to derive new upper bounds on the allowed diphoton branching ratio in the mass range $65 - 160$~GeV.

We found upper limits on branching ratios between $\sim 2 - 50\%$ depending on the mass and custodial representation (see~\fref{HBR}).~We have also re-derived constraints on the mass of a light fermiophobic Higgs boson ruling out masses below $\sim 110$~GeV if their coupling to photons is dominated by $W$ boson loops and they do not possess decays to BSM particles.~Given current constraints on the 125~GeV Higgs boson couplings, these bounds are already stronger than those obtained assuming SM-like production mechanisms~\cite{Abreu:2001ib,Abbiendi:2002yc,Heister:2002ub,Achard:2002jh,Abbott:1998vv,Affolder:2001hx,Chatrchyan:2013sfs} and will only get increasingly so as time goes on without observing a deviation from SM-like Higgs boson couplings.~We have also noted that these limits can be improved upon if current 13 TeV LHC diphoton searches~\cite{CMS-PAS-HIG-17-013,Aaboud:2016tru} are updated to cover the currently `open' window between $110-200$~GeV.~We then examined two simple ways to evade these searches via loop cancellations and/or decays into an invisible sector.

First we studied what level cancellations would give a suppression of the effective couplings to photons sufficiently large to escape LHC diphoton limits.~We find $\mathcal{O}(1-50\%$) cancellations between $W$ boson loops and other charged particles are needed.~We then explored two explicit scenarios in the Georgi-Machacek (GM) and supersymmetric GM (SGM) models which naturally contain custodial fermiophobic Higgs bosons.~In the case of the SGM there is a also a neutralino sector which opens up potential invisible decays that can drastically suppress the branching ratio into photons.~This leads to a significantly larger allowed parameter space found in the SGM model than in the GM model.~A study of the (custodial) superpartner fermion sector and examining LHC searches for gaugino/higgsinos with compressed spectra as well as potential dark matter phenomenology is ongoing~\cite{followup2}.

Finally, we examined the recently observed 95~GeV CMS diphoton excess, which has also been explored in various recent studies~\cite{Haisch:2017gql,Mariotti:2017vtv,Richard:2017kot}.~We have shown that for a custodial fiveplet Higgs boson, branching ratios $\sim 10\%$ are needed to explain the excess.~We found that this can be achieved with the custodial fiveplet present in the GM and SGM models if there is sufficient destructive interference between the $W$ boson loop and other (doubly) charged particles to suppress the diphoton branching ratio.~In the case of the SGM, a $\sim 95$~GeV diphoton signal may also imply a neutralino around $45 - 50$~GeV which could be targeted in LHC invisible searches

Extended Higgs sectors possessing custodial symmetry and fermiophobia with SM fermions can evade many of the experimental constraints which otherwise apply to extended Higgs sectors.~We encourage LHC experimental searches to utilize the Drell-Yan Higgs pair production plus diphoton searches emphasized here to shine light on these potentially hidden extended Higgs sectors.
%

~\\
\noindent
{\bf Acknowledgements:}~We thank Andrew Akeroyd, Filippo Sala, Jose Santiago, Daniel Stolarski, and Lorenzo Ubaldi for useful comments and discussions.~The work of R.V.M.~is supported by MINECO,~FPA 2016-78220-C3-1-P,~FPA 2013-47836-C3-2/3-P (including ERDF), and the Juan de la Cierva program,~as well as by Junta de Andalucia Project FQM-101.~The work of R.V.~is partially supported by the Sam Taylor fellowship.~K.X. is supported by U. S. Department of Energy under Grant No. DE-SC0010129.~K.X.~also thanks Fermilab for their hospitality and partial support during this work.



\bibliographystyle{apsrev}
\bibliography{refs_lightHiggs}

\end{document}